\documentclass[traditabstract]{aa}
\usepackage{graphicx}
\usepackage{natbib}

\usepackage{amsmath}
\usepackage{txfonts}

\usepackage{hyperref} 
\hypersetup{
  colorlinks=true,
  pdfborder=2 2 0.,
  linkcolor=blue,
  anchorcolor=black,
  citecolor=blue,
  urlcolor=black,
}

\usepackage{booktabs}
\bibpunct{(}{)}{;}{a}{}{,}

\usepackage{subfig}

\begin{document}

  \title{The HST/ACS star formation history of the Tucana dwarf spheroidal galaxy: clues from the horizontal branch}
  
  \author{A. Savino\inst{1,2}, E. Tolstoy\inst{1}, M. Salaris\inst{3}, M. Monelli\inst{4,5} \and T.J.L. de Boer\inst{6}}

\institute{Kapteyn Astronomical Institute, University of Groningen, Postbus 800, 9700 AV Groningen, The Netherlands
            \and Zentrum f\"{u}r Astronomie der Universit\"{a}t Heidelberg, Astronomisches Rechen-Institut, M\"{o}nchhofstr. 12, 69120 Heidelberg, Germany,
            \email{A.savino@uni-heidelberg.de}
            \and Astrophysics Research Institute, 
           Liverpool John Moores University, 
           IC2, Liverpool Science Park, 
           146 Brownlow Hill, 
           Liverpool L3 5RF, UK 
            \and Instituto de Astrofísica de Canarias, C/Via Lactea, s/n, 38205 La Laguna, Tenerife, Spain
            \and Departamento de Astrofísica, Universidad de La Laguna, 38206 La Laguna, Tenerife, Spain
            \and Institute for Astronomy, University of Hawaii, 2680 Woodlawn Drive, Honolulu, HI 96822, USA
           }

 \abstract{
We report a new star formation history for the Tucana dwarf spheroidal galaxy, obtained from a new look at a deep HST/ACS colour-magnitude diagram. We combined information from the main sequence turn-off and the horizontal branch to resolve the ancient star formation rates on a finer temporal scale than previously possible. We show that Tucana experienced three major phases of star formation, two very close together at ancient times and the last one ending between 6 and 8 Gyr ago. We show that the three discrete clumps of stars on the horizontal branch are linked to the distinct episodes of star formation in Tucana. The spatial distribution of the clumps reveals that each generation of stars presents a higher concentration than the previous one. The simultaneous modelling of the horizontal branch and the main sequence turn-off also allows us to measure the amount of mass lost by red giant branch stars in Tucana with unprecedented precision, confirming dwarf spheroidals to be excellent laboratories to study the advanced evolution of low-mass stars. 
}
\authorrunning{A. Savino et al.}
\titlerunning{The star formation history of the Tucana dSph}
  \maketitle


\section{Introduction}

Resolved stellar populations provide a valuable tool to peer into the past of nearby stellar systems. The stellar colour-magnitude diagrams (CMDs), and the ability to obtain spectra for individual stars, give us the opportunity to reconstruct very detailed star formation histories (SFHs) for our closest neighbours, as well as to probe their chemical enrichment and internal dynamics \citep[see, e.g.,][and references therein]{Mateo98, Gallart05,Tolstoy09}. While the Local Group is home to only three massive galaxies, smaller systems are much more numerous, allowing comparative studies for a substantial sample of galaxies.

One of the discoveries coming from the observations of Local Group dwarfs is that even very simple systems such as dwarf spheroidal galaxies (dSphs) host complex populations of stars, that range in metallicity, kinematics and spatial distribution. Aside from the very complex stellar populations of galaxies like Carina and Fornax \citep[e.g.,][]{Smecker94,Beauchamp95,Stetson98}, this has been detected also in simpler dSphs, such as Sculptor and Sextans \citep[e.g.,][]{Majewski99,Bellazzini01,Tolstoy04} and seems to be common in low mass galaxies. While several scenarios have been put forward to explain this phenomenon, such as galaxy mergers \citep{Amorisco12a,delPino15}, supernova feedback \citep[e.g.][]{Salvadori08,Revaz09}, and tidal interaction with larger galaxies \citep{Pasetto11}, the origin of these complexities is still unclear. Certainly,  an understanding of the processes that leave this imprint in the stellar population of dwarf galaxies would be a significant step forward to explain galactic formation and evolution.

\begin{figure}
\centering
\includegraphics[width=0.4\textwidth]{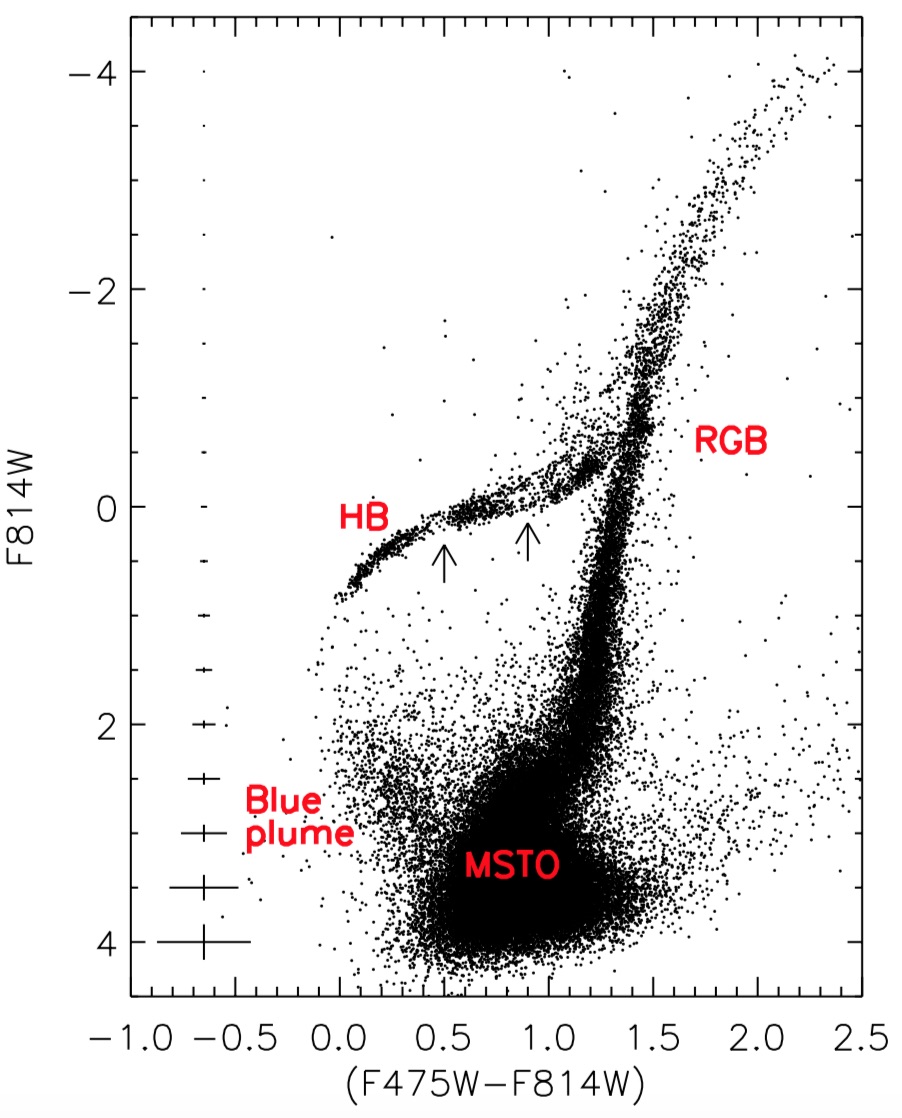}
\caption{The HST/ACS CMD of the Tucana dSph, corrected for distance and dust extinction, from \citet{Monelli10a}. The morphology of the HB is corrected for RR Lyrae variability. The major CMD features are indicated. The arrows mark the position of the HB gaps. The crosses on the left-hand side illustrate the average photometric uncertainty as a function of magnitude.}
\label{fig:CMD}
\end{figure}

Similarly to other nearby dSphs, the Tucana dSph has been found to host distinct stellar components \citep[e.g., ][]{Bernard08,Monelli10c}. Unlike most of the Local Group dSphs, which are satellites of M31 and the Milky Way, Tucana currently resides at the periphery of the Local Group, at a distance of $\sim 890$ kpc \citep{Bernard09}. This means that this galaxy has spent at least several Gyrs away from the enviromental disturbance of a large galaxy, but it is likely to have been a satellite of the Milky Way until $\sim 10$ Gyr ago \citep{Fraternali09,Teyssier12}.

The distance of Tucana from the Milky Way is also a complication when it comes to its observational characterisation, requiring photometric studies to be performed with high resolution instruments such as the Hubble Space Telescope (HST), as it has been done as part of the ``Local cosmology from isolated dwarfs" (LCID) project \citep[][hereafter M10. See Fig.~\ref{fig:CMD}]{Monelli10a}. In this paper we look again at these beautiful data taking advantage of a new method \citep{Savino18a} to model the CMD of Tucana including the horizontal branch (HB). This approach makes use of the simultaneous modelling of all the features observed in the CMD, to measure the SFH with improved resolution. Thanks to quantitative models of the HB, we obtain a detailed measurement of the early star formation in this galaxy, resolving distinct phases of activity. 
 
 \section{Dataset}
 \label{data}
 
The photometric data we use in this analysis have been acquired with the ACS camera, on board of the HST, as part of the LCID project  and have been presented in M10. The dataset consists of deep exposures in the $F475W$ and $F814W$ passbands, covering a field of view of $\sim3'.4 \times \sim3'.4$ \citep[for reference, the tidal radius of Tucana is $\sim3'.7$,][]{Mateo98}. We complement these data with the catalogue of RR Lyrae from \citet{Bernard09}, derived from the same LCID observations. When modelling the HB, it is critical to remove the effect of the variability of the stars in the instability strip. To this end, we cross-matched the positions of the detected RR Lyrae with the photometric catalogue, substituting to their observed magnitudes their intensity-averaged magnitudes. This is very close to the ``static" magnitude that these stars would have if they were not pulsating \citep{Bono95}. In this way, we are able to reconstruct the intrinsic morphology of the HB. To allow the comparison between models and observations, we transform the photometry into the absolute photometric reference frame, using the distance modulus $(m-M)_0 = 24.74$ and the extinction $A_V = 0.094$ \citep{Bernard09}.

The resultant $(F475W - F814W)$ vs $F814W$ CMD is shown in Fig.~\ref{fig:CMD}, where the superb quality of HST photometry can be appreciated. Despite the distance of this galaxy, stars are resolved down to $\sim 1.5$ magnitudes below the oldest main sequence turn-off (MSTO). Among the features that can easily be identified, are a prominent HB and a plume of bright blue stars emerging from the MSTO. The HB clearly has a complex structure. This was noted already by \citet{Harbeck01}, but it can be better appreciated with the increased photometric precision of this dataset. The distribution of stars across the HB is not uniform, but has three distinct clumps of stars. These are separated by lower stellar density ``gaps" as indicated in Fig.~\ref{fig:CMD} by arrows. The HB stars to the red of the reddest gap also show a bifurcation in magnitude. This is caused by stars that are evolving towards the asymptotic giant branch from the blue HB, following a brighter trajectory than red HB stars that are still in the early phases of helium-burning.

   \begin{figure}
\centering
\includegraphics[width=0.5\textwidth]{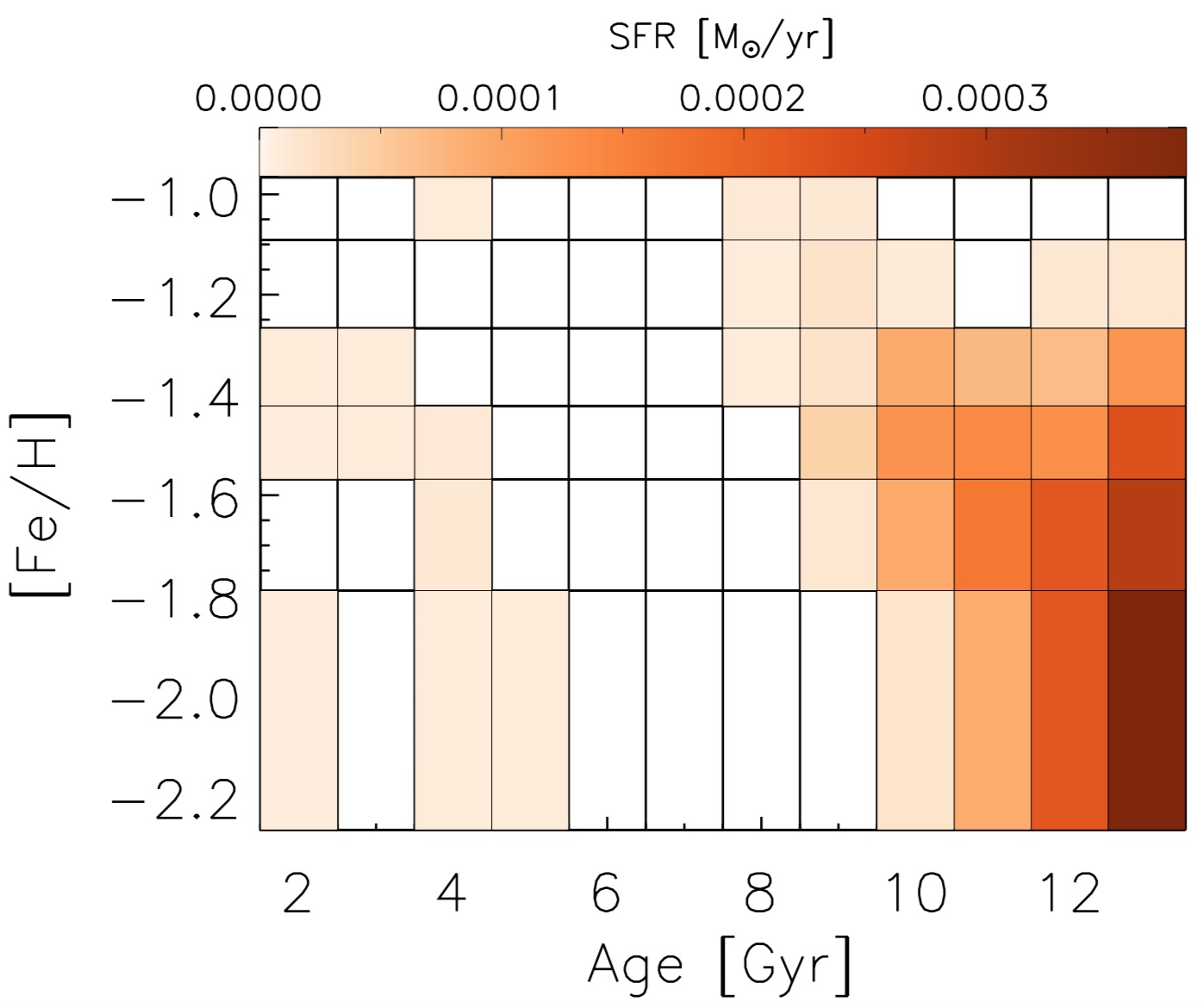}
\caption{Star formation rate distribution of Tucana, in the age-metallicity plane, according to the SFH measurement of \citet{Monelli10a}.}
\label{fig:Monelli}
\end{figure}

 \section{Interpreting the HB of Tucana}
 \label{HB}
A logical starting point for our analysis is to compare how well the properties of Tucana's HB match what we already know about this galaxy. Tucana has already been subject to a detailed photometric SFH analysis (M10) using the MSTO and the red giant branch (RGB).  This SFH is given in Fig.~\ref{fig:Monelli} and shows that Tucana is mainly composed of old stars, with the star formation terminating between 8 and 10 Gyr ago. Fig.~\ref{fig:Monelli} also contains a significant amount of star formation happening at ages younger than 6 Gyr and low metallicities. This component, generated by the fit of the blue plume above the MSTO, seems to defy the age-metallicity evolution expected from galaxy evolution and it is generally attributed to a population of blue stragglers, rather than to genuine star formation. Indeed, \citet{Monelli12} carried out an extensive analysis of blue plume stars in Tucana, concluding that they likely result from binary mass transfer, in accordance with what found in other dSphs \citep[e.g.][]{Momany07,Mapelli07}.

As first step, we will adopt the M10 SFH to predict the morphology of Tucana's HB. From the SFH of M10, we generated a synthetic CMD, modelling the evolution up to the tip of the RGB with BaSTI isochrones \citep{basti,Pietrinferni06}. The stellar phases after the tip of the RGB are modelled using a grid of theoretical HB tracks, from the same group, covering a mass range of $0.5-1.5 M_{\odot}$. This ensures that we are able to model even the reddest and most massive HB stars of Tucana, as well as the progeny of the blue straggler population. We adopt the RGB mass loss prescription from \citet{Salaris13}.  

Although the BaSTI theoretical library has been recently updated \citep{Hidalgo18}, we chose to use the original models computed with the ``canonical'' set-up. This is because the new BaSTI release is still limited to the scaled-solar chemical abundance mixture, while $\alpha$-enhanced tracks, more appropriate to model Tucana's old population, are currently being calculated (Hidalgo et al., in prep.). For the HB tracks, the main differences between the original and the new library are updated conductive opacities in the degenerate core, an updated abundance pattern for the solar mixture and  a different scaling between helium abundance and metallicity. In the $(F475W - F814W)$ vs $F814W$ CMD, this causes the new HB tracks to be slightly fainter (within 0.1 mags) and slightly bluer (within 0.05 mags) at fixed [Fe/H]. This difference is comparable to other sources of uncertainty in our modelling, such as the distance modulus determination or the precision of photometric colour indices on the HB,  and it corresponds to 1-2 bins when we convert the CMD to density maps (see \S~\ref{setup}).
 \subsection{The extension of the red HB}
 \label{redHB}

First we explore the colour extension of the HB and how it is linked to the end of the star formation in Tucana. Before we proceed we deem it useful to remind the reader how the parameters of a stellar population determine the temperature distribution on the zero-age HB (defined as the set of stellar model properties at the beginning of the helium-burning phase). Stars belonging to old stellar populations ignite core helium burning in lower initial masses compared to younger counterparts with the same chemical composition, and appear bluer on the HB. On the other hand, at fixed age, increasing the metallicity increases the mass of HB stars, moving them to progressively redder colours. In case of chemical self-enrichment typical of galaxy evolution, younger populations also tend to be on average more metal rich, hence they will be redder than the older, more metal poor, component. This straightforward picture is however complicated by the effect of RGB mass loss. Increasing the mass lost along the RGB will result in a lower HB mass for a given population age and chemical composition, mimicking the properties of an older and/or more metal poor population.  Knowing this quantity is thus fundamental for dating stellar populations through the HB.

  \begin{figure}
\centering
\includegraphics[width=0.5\textwidth]{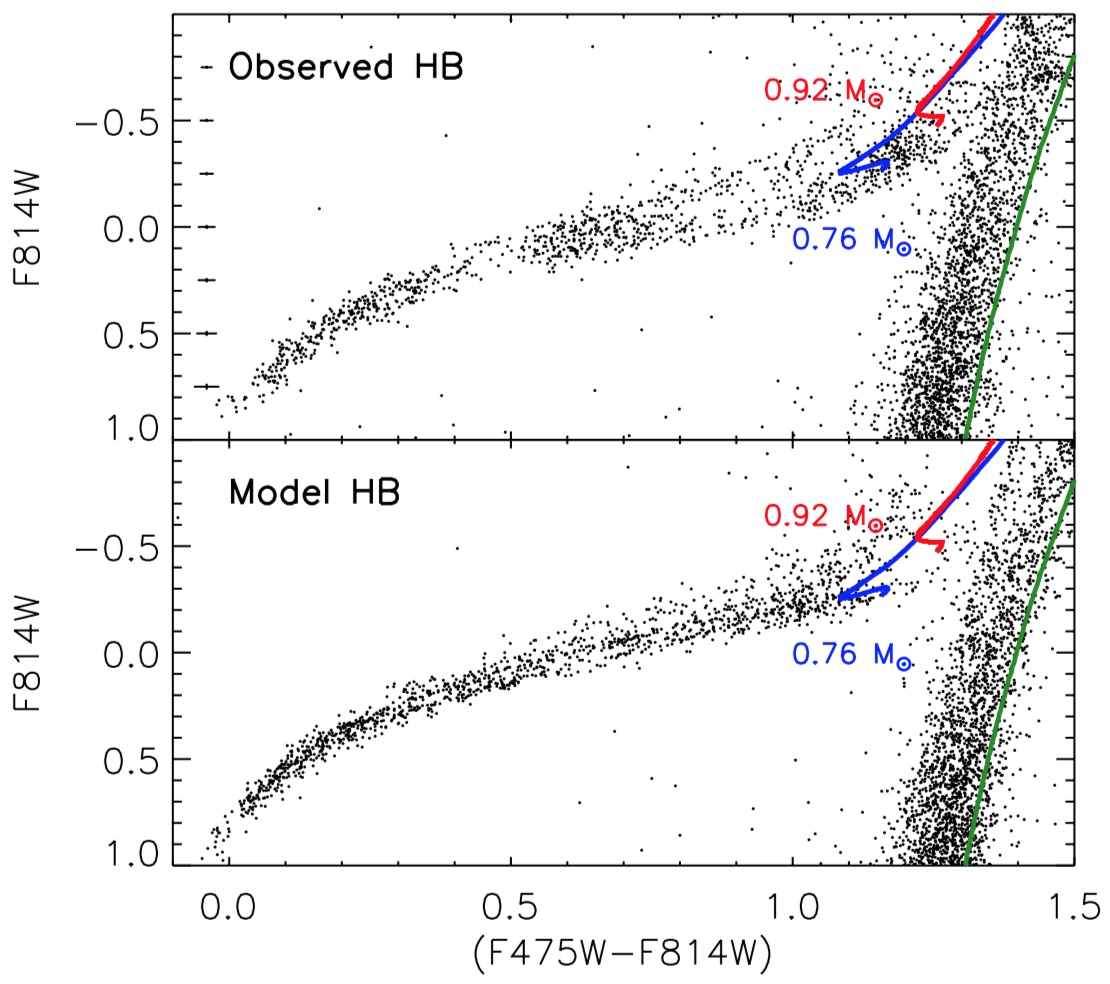}
\caption{ \textit{Upper panel:} Observed HB and RGB of Tucana. The crosses on the left-hand side illustrate the average photometric uncertainty as a function of magnitude \textit{Lower panel:} Synthetic CMD obtained with the SFH of \citet{Monelli10a}. A theoretical isochrone for t= 8 Gyr and [Fe/H] = -1.27, which has a mass at the RGB tip of $0.92 M_{\odot}$, is superimposed (green). The blue track represents the evolution of HB stars coming from this stellar population and experiencing RGB mass loss as prescribed by \citet{Salaris13}. The red track shows the evolution of the same stars, but with no RGB mass loss.}
\label{fig:ZAHB}
\end{figure}

 \begin{figure}
        \subfloat[][]
	{\includegraphics[width=0.4\textwidth]{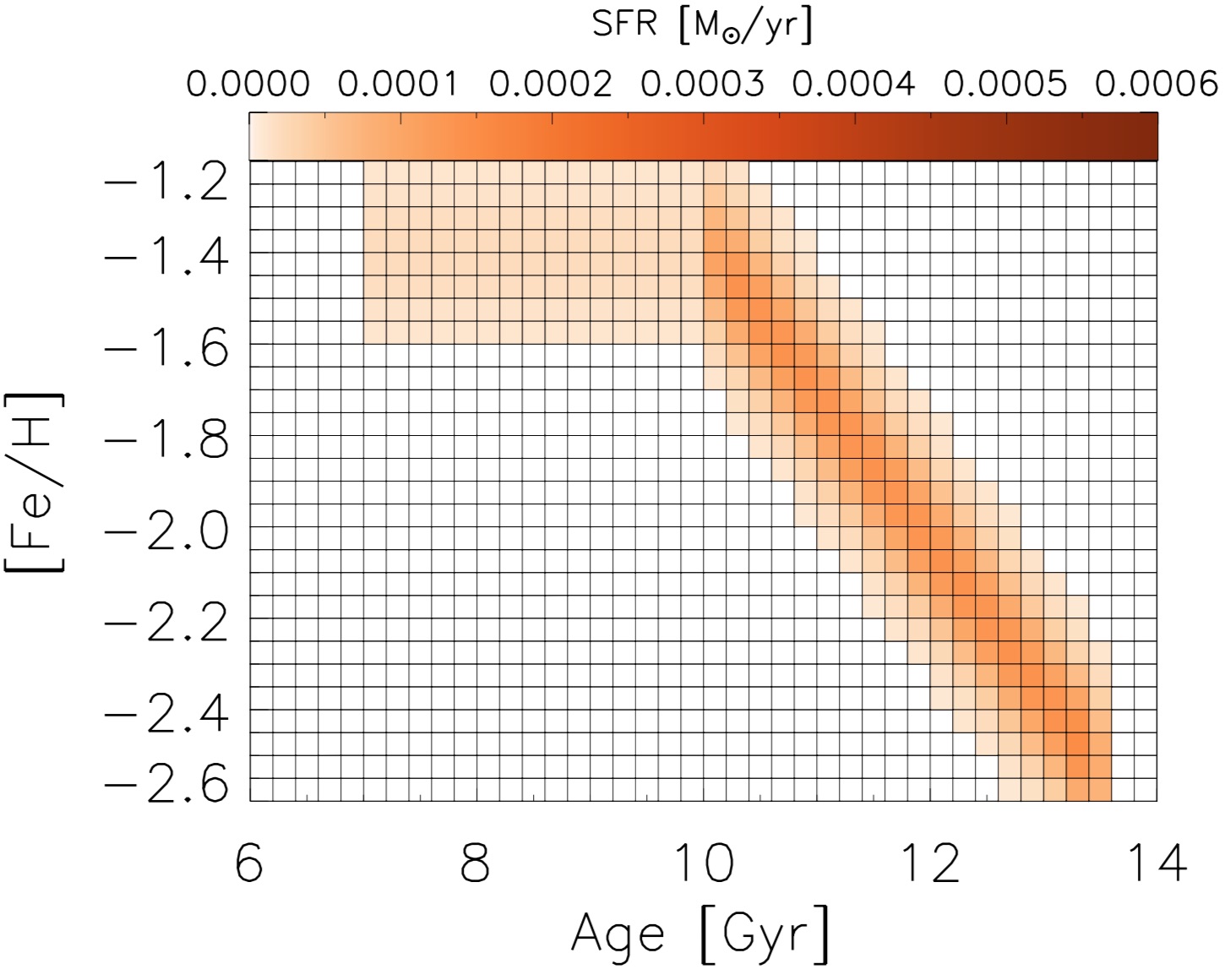} \label{fig:smooth}} \quad
	 \subfloat[][]
	{\includegraphics[width=0.4\textwidth]{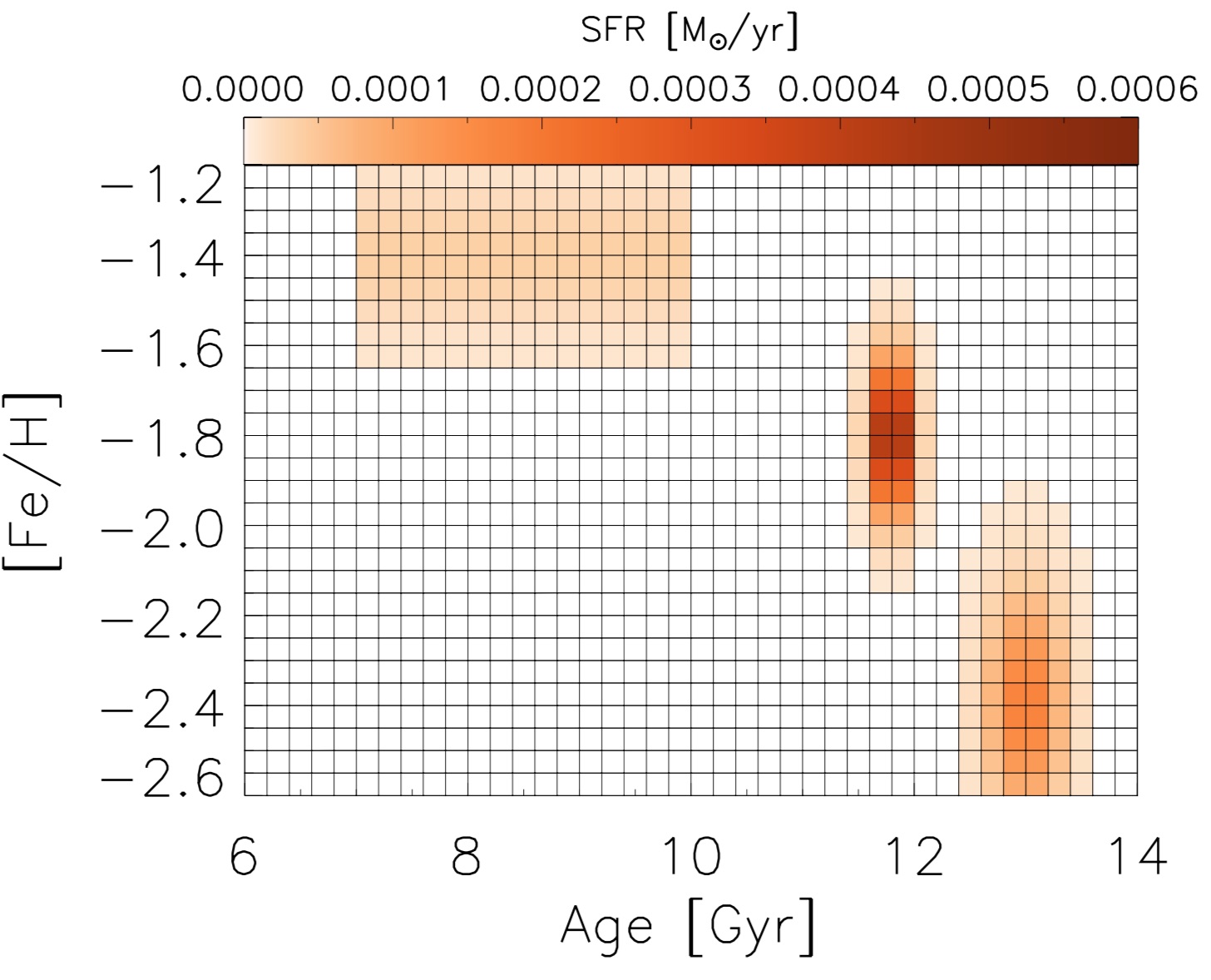} \label{fig:clumpy}} \quad
	
    \caption{Mock SFHs used to explore the origin of the HB gaps. a): Star formation distribution in the age-metallicity plane for the prolonged SFH. b) Star formation distribution in the age-metallicity plane for the bursty SFH.} 
    \label{fig:rad}
\end{figure}

\begin{figure*}
\centering
\includegraphics[width=\textwidth]{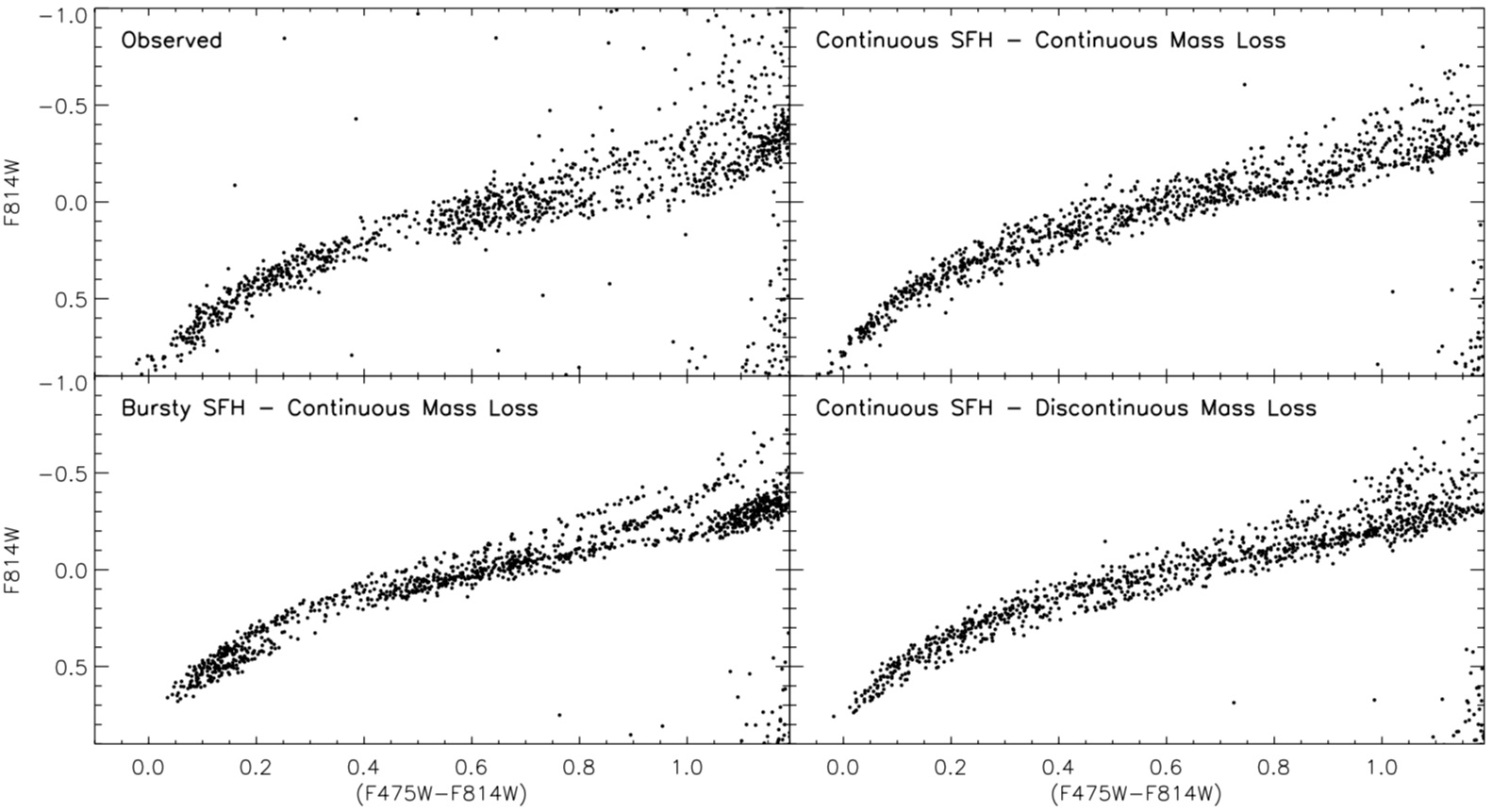}
\caption{The observed HB of Tucana (upper left), along with the synthetic HB obtained with different SFHs and mass loss prescriptions.}
\label{fig:HBcomp}
\end{figure*}

The amount of mass lost at high metallicities is strongly connected to the extension of the HB at the red end and it is a critical factor in determining when the star formation ceased in Tucana. Fig.~\ref{fig:ZAHB} shows the observed HB and RGB of Tucana (upper panel) and our prediction obtained from M10's SFH (lower panel). Superimposed there is a BaSTI isochrone with age of 8 Gyr and $\rm [Fe/H] = -1.27$.  This stellar population is roughly the youngest and most metal rich to have formed in Tucana, according to M10. The isochrone nicely coincides with the red boundary of the RGB, confirming the reasonable upper limit in metallicity. For this stellar population, the initial mass of stars at the tip of the RGB is $0.92 \, M_{\odot}$. Also HB evolutionary tracks for the same metallicity are plotted. The blue track corresponds to a $0.76 \, M_{\odot}$ star. This is the resulting HB mass that is obtained, from this stellar population, with an RGB mass loss of  $0.16 \, M_{\odot}$, in accordance with our mass loss prescription. The red track is for a $0.92 \, M_{\odot}$ star, which corresponds to no RGB mass loss at all. It can be seen from Fig.~\ref{fig:ZAHB} that the extent of the red HB in Tucana is compatible with the SFH from M10 only if no mass is lost on the RGB. This is not a realistic assumption. Assuming a non-zero mass loss value implies that most of the stars on the red HB have formed at more recent times. The higher the mass loss value, the more recently the star formation stopped.

 \subsection{The origin of the HB gaps}
 \label{gaps}

Another interesting difference is that our model HB does not show the two prominent gaps seen on the HB of Tucana. These features have been observed in several Local Group dSph galaxies and seem to be relatively common \citep[e.g.][]{Savino15}. Gaps were also detected on the HB of Galactic globular clusters \citep[e.g.][]{Catelan98,Ferraro98,Dcruz00}. However, the parameters governing the HB morphology have different distributions in globular clusters and in dSphs, therefore it is unlikely that gaps in the two classes of objects share a common origin. Nevertheless, we can take advantage of the tools developed to analyse the statistical significance of the HB gaps in globular clusters and apply them to Tucana.

There is, in fact, the possibility that the drop in stellar counts its simply due to stochasticity. Following the procedure laid out in \citet{Catelan98} we calculate the probability that sampling a uniform stellar distribution along the HB would result in underpopulated regions as observed in Tucana. We found this probability to be $6.95\cdot 10^{-4}$ for the redder gap and $2.05\cdot 10^{-14}$ for the bluer gap. Such low probabilities are due to the high number of stars on the HB of Tucana and suggest that the gaps are real features.

As already mentioned, the age and metallicity evolution of galaxy stellar populations tend to move the colour of helium-burning stars from the blue to the red. The presence of gaps in the colour distribution of the HB could thus be interpreted as discreteness in the age and metallicity distribution of stellar populations in the galaxy. However, the morphology of the HB is determined by the combination of SFH and RGB mass loss. We thus explore whether the presence of clumps on the HB is necessarily caused by distinct stellar populations or if it can arise because of sharp variations in the RGB mass loss as a function of stellar parameters. We investigate this question by examining the morphology of synthetic HB models, calculated with different assumptions on the shape of the SFH and the mass loss function. We examine three cases. The first is a prolonged event of star formation (Fig.~\ref{fig:smooth}), that roughly matches the age and metallicity range of Tucana. In this first case, the stars lose mass according to a smoothly increasing function of metallicity, following the relation presented in \S~\ref{ML}. The second case uses the same SFH, but with a discontinuous, step-function, mass loss that obeys the relation:

 \begin{figure*}
\centering
\includegraphics[width=\textwidth]{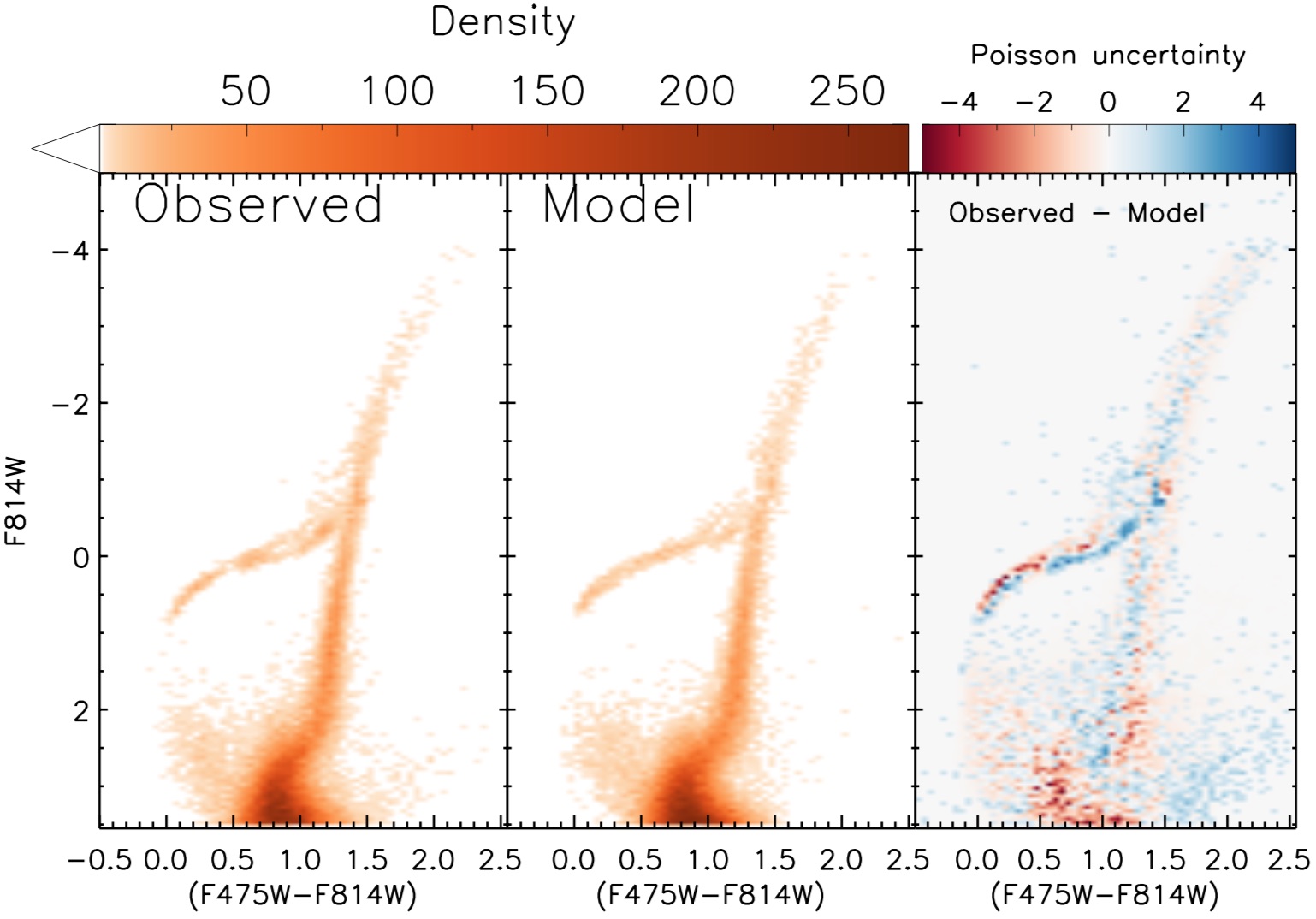}
\caption{Observed (left) and best-fit (centre) Hess diagrams for Tucana, colour coded by stellar density. The right panel shows the stellar count residuals, expressed in terms of the Poisson error.}
\label{fig:Fit}
\end{figure*}

\begin{equation}
\Delta M_{RGB} = 
\begin{cases}
\rm 0.08\, M_{\odot}   \quad if  &\rm [Fe/H] < -2.1,\\
\rm 0.14\, M_{\odot}  \quad if  \quad -2.1\leq\!\! & \rm [Fe/H] < -1.55 ,\\
\rm 0.20\, M_{\odot}  \quad if &\rm [Fe/H] \geq -1.55.
\end{cases}
\end{equation}
These values have been selected to match the average of the smooth mass loss relation over the same metallicity interval. The third case is calculated with the smooth mass loss function and with a SFH composed of three distinct star formation events (Fig.~\ref{fig:clumpy}).

The synthetic HBs from the three set-ups are shown in Fig.~\ref{fig:HBcomp}, along with the observed HB of Tucana. A continuous SFH and a continuous mass loss predictably result in a uniformly populated HB. The interesting result is that having discontinuities in the mass loss is not sufficient to create gaps in the HB. The only way of creating these features is to have a complex SFH, composed of distinct star formation events. The inefficiency of mass loss in creating gaps in the HB stellar distribution was already suggested by \citet{Savino15}, analysing the Carina dSph, and it is further supported by this experiment, indicating that a clumpy HB is most likely due to discontinuous star formation episodes.

 \section{Global modelling of Tucana's CMD}
 \label{Model}
 The work laid out in the previous section has highlighted how the features present on the HB contain a wealth of information that can be exploited to obtain a more detailed SFH than from the MSTO/RGB alone. In this section we present a global modelling of the CMD of Tucana, that combines the modelling of the HB with that of other evolutionary features, such as the MSTO, to provide an updated SFH for the Tucana dSph.
 
 \subsection{The set-up}
 \label{setup}
 
We model the CMD using {\small MORGOTH} \citep{Savino18a}. {\small MORGOTH} is a SFH recovery method that uses synthetic CMDs to model all the evolutionary phases up to the beginning of the thermally pulsating asymptotic giant branch. This permits us to exploit the information contained in the HB. As the morphology of the HB is dependent on the RGB mass loss, this quantity is left as a free parameter that is measured along with the SFH. The total amount of mass lost on the RGB by a given star is assumed to have a linear dependence on metallicity. The procedure {\small MORGOTH} uses to treat the HB model generation ensures that the number counts and stellar distribution across the CMD are always consistent with the mass loss adopted. We refer the reader to \citet{Savino18a} for more details.

To generate synthetic CMD models we need several ingredients. As mentioned in \S~\ref{HB}, we use stellar isochrones and HB evolutionary tracks from the BaSTI theoretical library. We generate the synthetic models using a Kroupa initial mass function \citep{Kroupa01} and a binary fraction of 0.4, as done in M10. The models are calculated covering ages from 1 to 14 Gyr, with a bin size of 0.5 Gyr, and values of [Fe/H]  from --2.6 to --0.6, with a bin size of 0.2 dex. The CMDs are transformed into density maps (Hess diagrams) with a bin size of 0.05 mags in both colour and magnitude. We exclude the faintest 0.5 magnitudes from the fit. This region is well below the MSTO and is heavily affected by incompleteness and photometric uncertainties, which are difficult to model in this low signal to noise region.

Due to the distance of Tucana, no direct measurements have been possible to determine the detailed chemical properties of its stars. For old populations, stars with the same global metal abundance [M/H], but different values of $\rm [\alpha/Fe]$, will have very similar structures \citep{Salaris93}. However, the opacity of the stellar atmosphere is much more sensitive to the chemical pattern \citep{Cassisi04}, especially in the blue region of the spectrum, thus affecting the observed magnitudes and colours. It is thus important to make a realistic assumption about the $\alpha$-enhancement profile of Tucana. We expect the oldest stars to be $\alpha$-enhanced, as observed in the Milky Way halo and in many old stellar systems. However, younger and more metal rich stars, are likely to have a scaled-solar mixture or even be $\alpha$-depleted \citep{Tolstoy09}. The metallicity at which $\rm [\alpha/Fe]$ starts to decrease and the slope of its relation with [Fe/H] depend on the chemical enrichment history of the galaxy. For our modelling we use the $\rm [\alpha/Fe]$ vs $\rm [Fe/H]$ profile of the Sculptor dSph, from spectroscopic measurements \citep{Battaglia08,Tolstoy09}. Broadly speaking, Sculptor has similar stellar mass and formation history to Tucana, motivating our assumption. We assume all stars with $\rm [Fe/H] < -1.84$ to have $\rm [\alpha/Fe] = 0.4$. For higher [Fe/H], the alpha enhancement decreases with $\rm \frac{d[\alpha/Fe]}{d[Fe/H]} = -0.64$, down to a minimum value of $\rm [\alpha/Fe] = -0.2$. In section~\ref{SFH} we discuss the impact of adopting a different alpha enhancement profile.

Finally, we need to model photometric uncertainties and incompleteness. This is done by means of artificial star tests, which we take from M10. These tests are made by injecting synthetic stars into the observation images and measuring their photometric properties, and they tell us the photometric error distribution and the completeness level as a function of position in the CMD. We use this information to include observational effects in the synthetic CMDs.

 \subsection{The best-fit model CMD}
 
 We run {\small MORGOTH} with the set-up described in the previous section and obtain the best-fit CMD shown in Fig.~\ref{fig:Fit}. The agreement with the observed CMD, also shown, is good. The broad morphology of all the stellar evolutionary phases is well reconstructed, including the colour extension of the HB. From the analysis of the residuals, approximately $64\%$ of the Hess bins where at least one star is observed have their star counts reproduced within one Poisson uncertainty and approximately $98\%$ have it within three Poisson uncertainties.

However, certain regions of the CMD show structures in the residual map. A clear discrepancy between the model and the observations regards the position of the RGB bump (located around $F814W = 1$). Such difference is not particularly problematic as it is long known that theoretical RGB bump luminosities are overpredicted at the level of 0.2-0.4 magnitudes \citep{DiCecco10, Monelli10c, Cassisi11,Troisi11}. The other major differences regard the stellar distribution on the HB. Although our model HB is not uniformly populated, it is not as clumpy as the observed HB. This creates clear residuals at the position of the observed HB gaps, especially for the bluer gap. Additionally, we seem to underpredict slightly the stellar counts in the reddest part of the HB. The fact that we are unable to completely eliminate these problems on the HB is not too surprising. Both the HB and the MSTO are very sensitive to the SFH. As we are, for the first time, quantitatively combining the independent constraints from these two regions to obtain a self-consistent SFH, some degree of disagreement is to be expected. However, the global morphology of the HB still has a significant influence on the fit and on the features of the new SFH. Furthermore, we can use the results of \S~\ref{HB} to evaluate the impact of these discrepancies on the interpretation of the SFH we recover.

\subsection{The RGB mass loss}
\label{ML}
The simultaneous modelling of multiple, age sensitive, CMD features breaks the degeneracies on the HB and allows us to measure the RGB mass loss parameters that best reproduce the observations. From our full {\small MORGOTH} CMD modelling we recover a mass loss that follows the relation:
\begin{equation}
 \rm \Delta M_{RGB} = [Fe/H]\times (0.089 \pm 0.014) + (0.300 \pm  0.025) \, M_{\odot}
\label{eq:ML}
\end{equation}
with a correlation between the slope and zero-point of the relation of 0.995.

\begin{figure}
        \subfloat[][]{\raisebox{-50.5ex}
	{\includegraphics[width=0.5\textwidth]{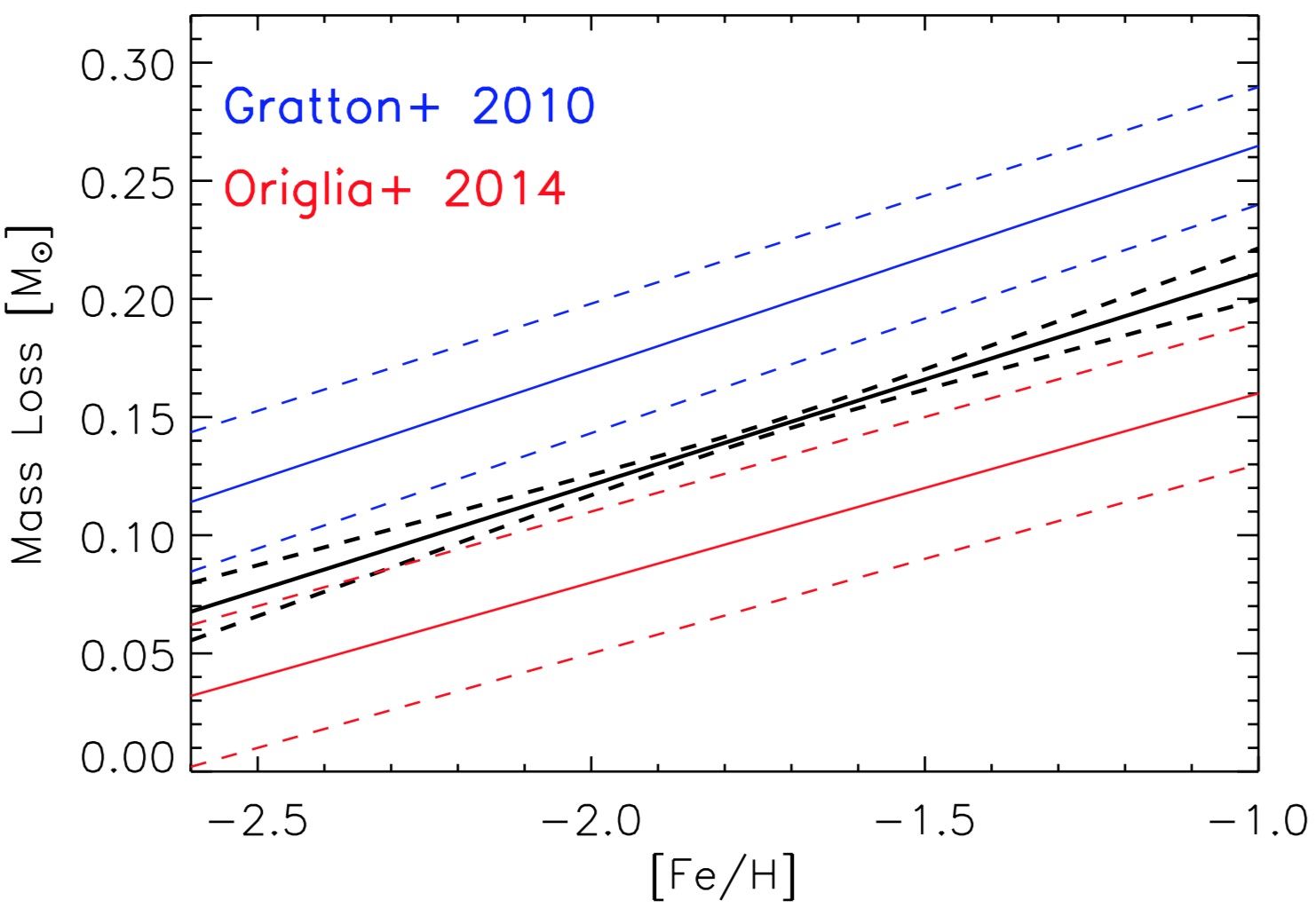}}} \quad
	 \subfloat[][]{\raisebox{-47ex}
	{\includegraphics[width=0.5\textwidth]{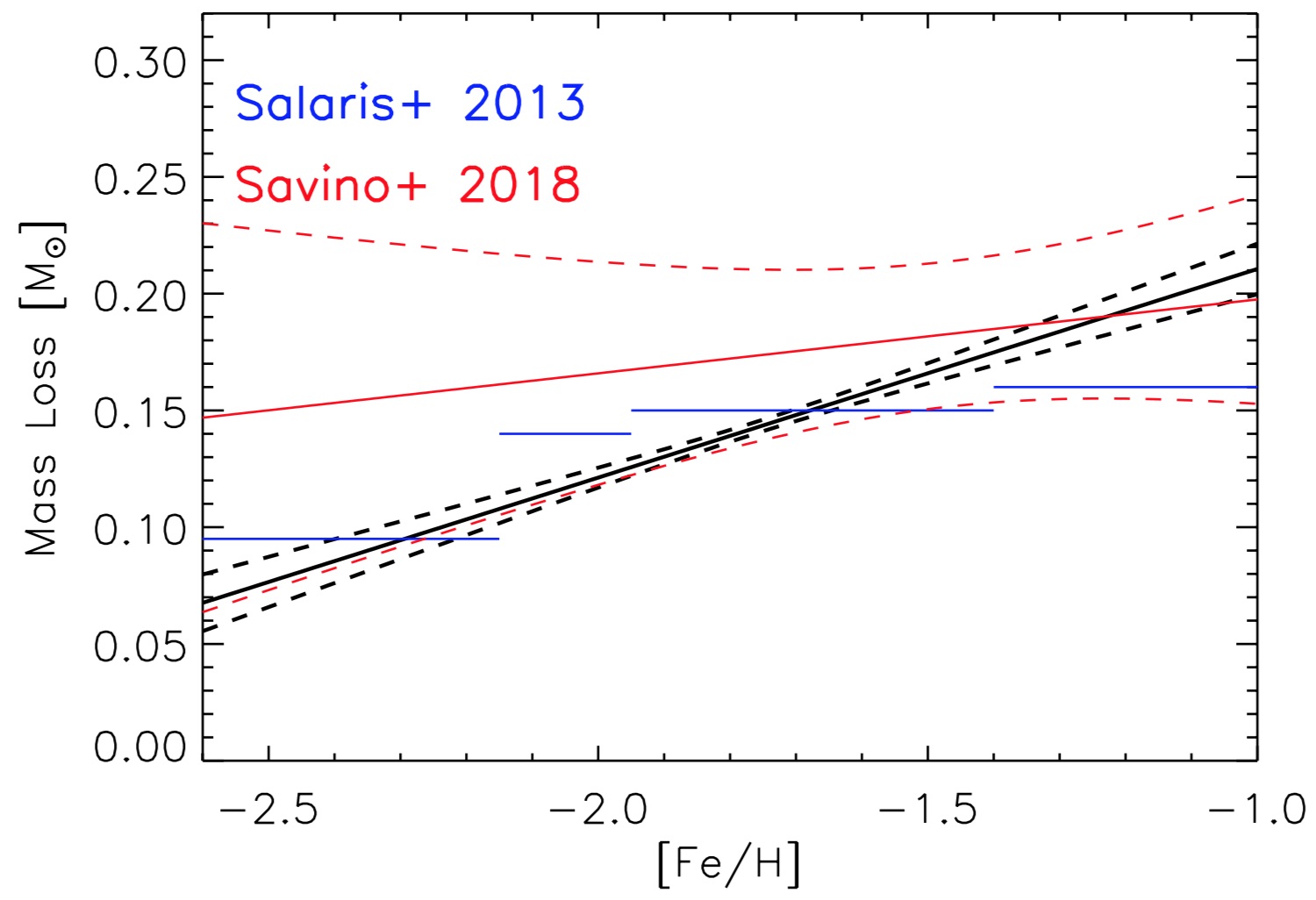} \label{fig:MLscl}}} \quad
    \caption{Total RGB mass loss as a function of metallicity, as measured from our modelling. The black solid line shows our measurements and the dashed black lines mark the one sigma confidence interval. a) Comparison with measurements obtained, on globular clusters, by \citet{Gratton10} (blue), \citet{Origlia14} (red). b) Comparison with measurements obtained, on the Sculptor dSph, by \citet{Salaris13} (blue) and \citet{Savino18a} (red).}
    \label{fig:ML}
\end{figure}

The RGB mass loss of our model is shown in Fig.~\ref{fig:ML}, along with other measurements from the literature, derived from Galactic globular clusters and from the Sculptor dSph. Estimates for the total RGB mass loss in globular clusters were obtained by \citet{Gratton10} and \citet{Origlia14}. The systematic difference between these two studies can be ascribed to the completely independent methodologies used. \citet{Gratton10}  calculated the mass loss from the median colour of the HB. However, globular clusters are known to host a significant number of stars with enhanced helium abundance \citep[more than 50\% in most cases,][]{Milone17}. Helium-enhanced populations have bluer HB colours for a given total RGB mass loss and population age, because they have a lower MSTO mass. Using the HB median colour without accounting for the helium abundance thus results in a systematic overestimate of the RGB mass loss rate. In contrast, \citet{Origlia14} measured the infrared excess of RGB stars and this approach relies on several assumptions about the properties of the circumstellar material. In particular, the gas-to-dust ratio used results in a very conservative, low value, of the RGB mass loss. The mass loss relation in the Sculptor dSph has been obtained by \citet{Salaris13} and \citet{Savino18a} using an analogous approach to this work. The two measurements also show some differences. However, the large uncertainties can easily account for this.

In agreement with other studies, we find that stars of increasing metallicity lose more mass. Quantitatively, our measurement lies in between the results of \citet{Gratton10} and \citet{Origlia14}, obtained from Galactic globular clusters, and it is compatible with the measurement of \citet{Salaris13} and \citet{Savino18a} for the Sculptor dSph. In spite of the sizeable uncertainties on the slope and zero-point of our mass loss relation, the correlation between these two quantities makes the uncertainties on the total RGB mass loss small. The nominal uncertainty on the total mass loss is less than $0.003 M_{\odot}$ for [Fe/H] between --1.9 and --1.6, where most of Tucana's stars are. However, the mass sampling of our HB model grid is $0.005 M_{\odot}$, so this number is a more realistic lower limit for the measurement uncertainty. For stars outside this metallicity range, where we have less constraining power, the measurement error is smaller than $0.015 M_{\odot}$.

 \begin{figure}
        \subfloat[][]
	{\includegraphics[width=0.5\textwidth]{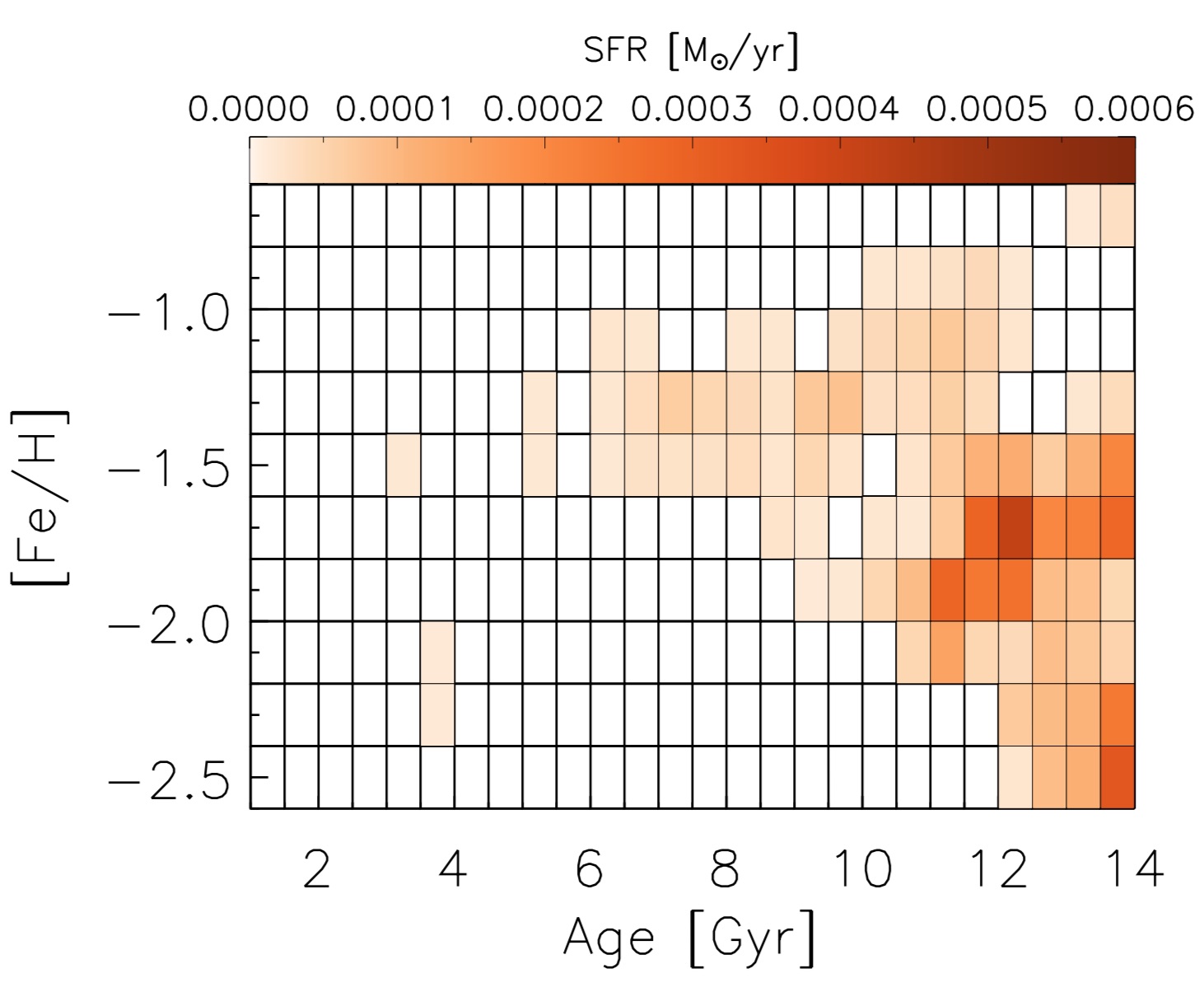} \label{fig:plane}} \quad
	 \subfloat[][]
	{\includegraphics[width=0.5\textwidth]{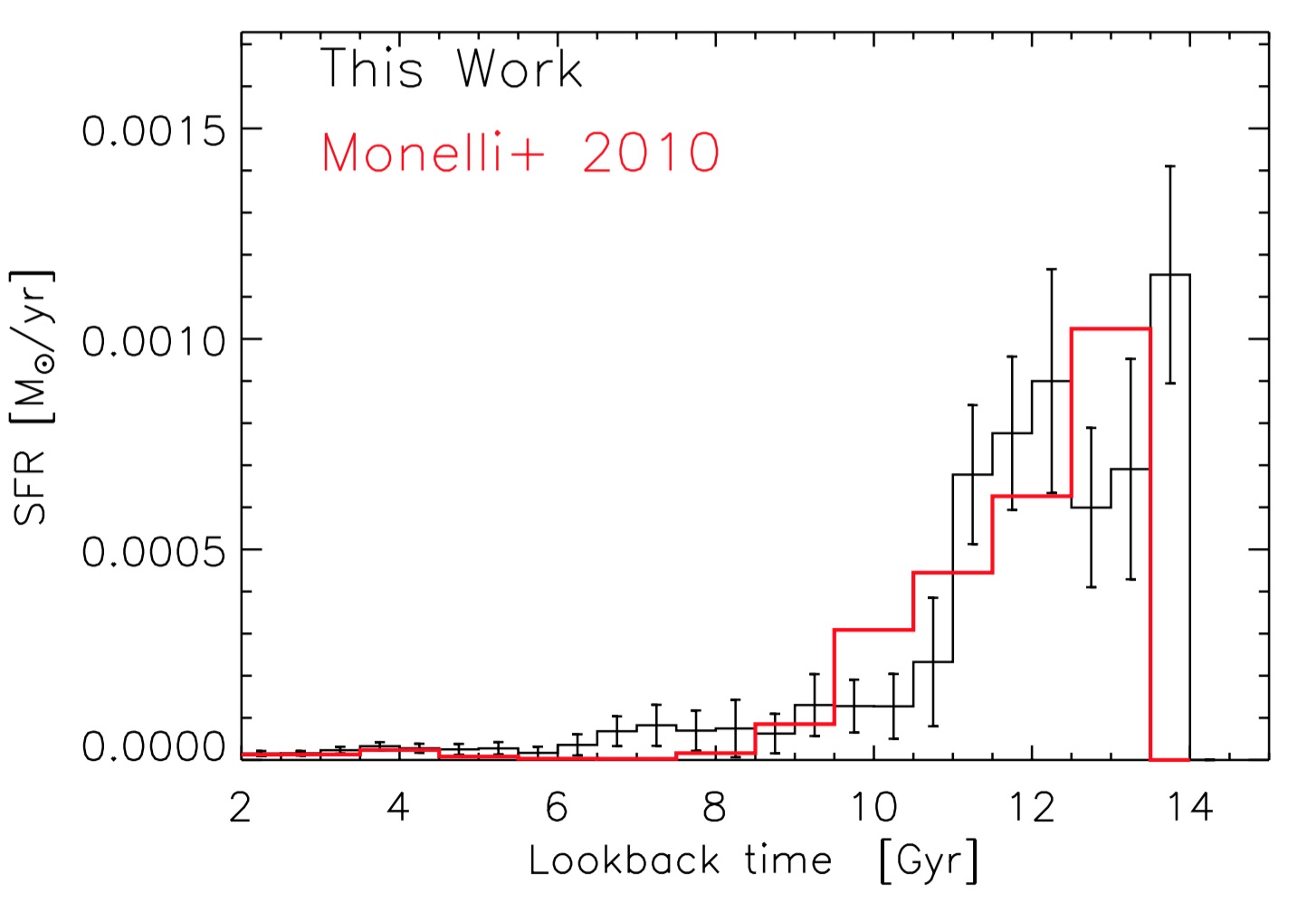} \label{fig:SFR}} \quad
    \caption{a) Best-fit SFH in the age-metallicity plane, colour coded by star formation rate. b) Corresponding star formation rate as a function of cosmic lookback time. The red histogram represents the best-fit SFH from \citet{Monelli10a}.}
    \label{fig:SFH}
\end{figure}
 
 \subsection{A new SFH for Tucana}
 \label{SFH}
 
 \begin{table}
\centering
\caption{The SFH corresponding to our best-fit model CMD. The table lists the mean [$\alpha$/Fe], [Fe/H] and age of the SFH bins, the corresponding star formation rate and its uncertainty. The bins have a width of 0.1 dex in [$\alpha$/Fe], 0.2 dex in [Fe/H] and 0.5 Gyr in age. The star formation rate is assumed to be constant in the bin. The complete table is available in the electronic version of the paper.}
\begin{tabular} {rcccc}
\toprule
$[\alpha$/Fe] & [Fe/H] &  Age &  SFR &  $\sigma_{SFR}$\\
 & & Gyr & $M_{\odot}$/yr & $M_{\odot}$/yr \\
 \midrule
0.40 & -2.5 & 1.25 & $1.27\cdot10^{-6}$ & $7.38\cdot10^{-7}$\\
0.40 & -2.3 & 1.25 & $2.58\cdot10^{-6}$ & $5.55\cdot10^{-7}$\\
0.40 & -2.1 & 1.25 & $2.09\cdot10^{-6}$ & $5.29\cdot10^{-7}$\\
0.40 & -1.9 & 1.25 & $1.97\cdot10^{-7}$ & $3.08\cdot10^{-7}$\\
0.31 & -1.7 & 1.25 & $3.31\cdot10^{-6}$ & $1.76\cdot10^{-6}$\\
0.18 & -1.5 & 1.25 & $1.95\cdot10^{-6}$ & $1.61\cdot10^{-6}$\\
0.05 & -1.3 & 1.25 & $9.91\cdot10^{-7}$ & $1.07\cdot10^{-6}$\\
-0.07 & -1.1 & 1.25 & $1.98\cdot10^{-7}$ & $5.04\cdot10^{-7}$\\
-0.20 & -0.9 & 1.25 & $1.08\cdot10^{-7}$ & $3.41\cdot10^{-7}$\\
-0.20 & -0.7 & 1.25 & $7.10\cdot10^{-7}$ & $9.39\cdot10^{-7}$\\
0.40 & -2.5 & 1.75 & $3.40\cdot10^{-7}$ & $3.66\cdot10^{-7}$\\
0.40 & -2.3 & 1.75 & $2.32\cdot10^{-7}$ & $3.91\cdot10^{-7}$\\
0.40 & -2.1 & 1.75 & $9.45\cdot10^{-7}$ & $1.01\cdot10^{-6}$\\
0.40 & -1.9 & 1.75 & $2.41\cdot10^{-7}$ & $4.37\cdot10^{-7}$\\
0.31 & -1.7 & 1.75 & $1.88\cdot10^{-7}$ & $4.89\cdot10^{-7}$\\
0.18 & -1.5 & 1.75 & $1.64\cdot10^{-6}$ & $2.17\cdot10^{-6}$\\
0.05 & -1.3 & 1.75 & $1.57\cdot10^{-6}$ & $1.94\cdot10^{-6}$\\
-0.07 & -1.1 & 1.75 & $3.21\cdot10^{-6}$ & $2.72\cdot10^{-6}$\\
0.20 & -0.9 & 1.75 & $1.42\cdot10^{-6}$ & $1.86\cdot10^{-6}$\\
0.20 & -0.7 & 1.75 & $1.30\cdot10^{-9}$ & $1.35\cdot10^{-8}$\\
...&...&...&...&...\\

\bottomrule
\end{tabular}
\label{tab:sfh}
\end{table}
 
Fig.~\ref{fig:SFH} shows the SFH corresponding to our best-fit CMD, in the age-metallicity plane (Fig.~\ref{fig:plane}), and the star formation rate as a function of cosmic lookback time (Fig.~\ref{fig:SFR}). The SFH is also reported in Table~\ref{tab:sfh}. We warn the reader that the individual star formation rates are highly correlated with each other and the listed uncertainties should not be used without the full covariance matrix. Fig.~\ref{fig:SFH} shows that the majority of Tucana's stars formed more than 11 Gyr ago, in line with M10. However, our model contains an extended tail of star formation, persisting until 6 Gyr ago. We stress that this younger and more metal rich population is not caused by the plume of blue stragglers. The ages that our model infers for these stars, in fact, are all younger than 5.5 Gyr, in accordance with M10. Rather, the more extended SFH is the result of the extension of the red HB. Integrating the SFH, with a Kroupa initial mass function \citep{Kroupa01}, results in a stellar mass (at formation) of $3.13 \pm 0.14  \cdot 10^6 M_{\odot}$ within the observed field of view.

The SFH is clearly not unimodal, but rather it is composed of three distinct star formation phases. The two stronger events occurred very early on and are separated by $\sim1$ Gyr, with the second being ~0.6 dex more metal rich than the first one. The last star formation event, of lower intensity, started about 10 Gyr ago and lasted for several Gyr, with metallicities as high as $\rm [Fe/H] = -1.0$.

The recovery of the SFH from the CMD analysis relies on a number of measurements and assumptions to create the synthetic stellar population models. For this reason it is important to perform a sanity check and verify the robustness of our results with respect to these inputs. To this end, we repeated the SFH measurement many times, varying parameters such as distance, reddening or binary fraction. We also used different $\rm [\alpha/Fe]$ vs [Fe/H] relations, varying the knee position and the slope of the relation. We also repeated the measurement assuming all stars to have either $\alpha$-enhanced ($\rm [\alpha/Fe] = 0.4$) or scaled-solar composition. The complete list of input set-ups is reported in Table~\ref{tab:tests}, where we quantify the change in the resulting SFH through the Kolmogorov-Smirnov distance:
\begin{equation}
D_{KS}=\sup{\mid F(t)-F^\prime(t) \mid}
\end{equation}
with $F(t)$ and $F^\prime(t)$ being the normalised cumulative distribution functions of the star formation rate, as function of time, in the nominal and modified set-up, respectively. $D_{KS}$ can assume values between 0 and 1, with 0 corresponding to identical distributions.

\begin{table}
\centering
\caption{Different set-ups employed to check the SFH dependence on the input parameters. The table lists the distance modulus {$(m-M)_0$}; the reddening E(B--V); the plateau, knee and slope values of the $\rm [\alpha/Fe]$ vs [Fe/H] relation; the binary fraction $f_{bin}$; and the Kolmogorv-Smirnov distance $D_{KS}$ of the resulting solution from the nominal solution.The first line shows the nominal set-up. For each line, entries in bold mark the parameters that deviate from the nominal set-up.}
\begin{tabular} {ccccccc}
\toprule
$(m-M)_0$ & E(B--V) &  Plateau &  Knee &  Slope  & $f_{bin}$ & $D_{KS}$\\
 \midrule
24.74 & 0.03& 0.4 & -1.84 & -0.64 & 0.4 &  0\\      
\bf24.70 & 0.03&0.4 & -1.84 & -0.64 & 0.4  & 0.03\\  
\bf24.72 & 0.03&0.4 & -1.84 & -0.64 & 0.4  & 0.02\\  
\bf24.76 & 0.03&0.4 & -1.84 & -0.64 & 0.4  & 0.02\\  
\bf24.78 & 0.03&0.4 & -1.84 & -0.64 & 0.4  & 0.05\\  
\bf24.80 & 0.03&0.4 & -1.84 & -0.64 & 0.4  & 0.06\\  
\bf24.82 & 0.03&0.4 & -1.84 & -0.64 & 0.4  & 0.10\\  
\bf24.84 & 0.03&0.4 & -1.84 & -0.64 & 0.4  & 0.09\\  
24.74 &\bf 0.00&0.4 & -1.84 & -0.64 & 0.4  & 0.04\\  
24.74 &\bf 0.01&0.4 & -1.84 & -0.64 & 0.4  & 0.05\\  
24.74 &\bf 0.02&0.4 & -1.84 & -0.64 & 0.4  & 0.03\\  
24.74 &\bf 0.04&0.4 & -1.84 & -0.64 & 0.4  & 0.10\\  
24.74 &\bf 0.05&0.4 & -1.84 & -0.64 & 0.4  & 0.05\\  
24.74 &\bf 0.06&0.4 & -1.84 & -0.64 & 0.4  & 0.09\\  
24.74 & 0.03&\bf0 &\bf - & \bf0 & 0.4 &  0.13\\  
24.74 & 0.03&0.4 & \bf -  & \bf 0 & 0.4  & 0.16\\  
24.74 & 0.03&0.4 &\bf -1.70 & -0.64 & 0.4  & 0.06\\  
24.74 & 0.03&0.4 & \bf-2.00 & -0.64 & 0.4  & 0.08\\  
24.74 & 0.03&0.4 & -1.84 &\bf -0.80 & 0.4  & 0.08\\  
24.74 & 0.03&0.4 & -1.84 &\bf -0.40 & 0.4  & 0.03\\  
24.74 & 0.03& 0.4 & -1.84 & -0.64&\bf 0  & 0.04\\  
24.74 & 0.03& 0.4 & -1.84 & -0.64&\bf0.2  & 0.04\\  
24.74 & 0.03& 0.4 & -1.84 & -0.64&\bf0.6  & 0.04\\  
 
\bottomrule
\end{tabular}
\label{tab:tests}
\end{table}

 Although these tests changed slightly the details of the SFH, none of the assumed set-ups significantly altered its shape. The strongest effect on the SFH, especially when changing the $\rm [\alpha/Fe]$ profile, is the ability to resolve the two oldest star formation events, which can become blended together. The tail of more recent and metal rich star formation is the most resilient to systematics in the modelling parameters, and it is always present in our SFH.  As for any other SFH measurement obtained from CMD modelling, the uncertainties on the parameters adopted to create the CMD models mainly affect the absolute age of the stellar populations, rather than the relative age distribution. A change in distance modulus of 0.05 mag, for instance, corresponds to a change in the age inferred from the MSTO of 0.5 $-$1 Gyr, depending on metallicity.

\begin{figure}
\centering
\includegraphics[width=0.5\textwidth]{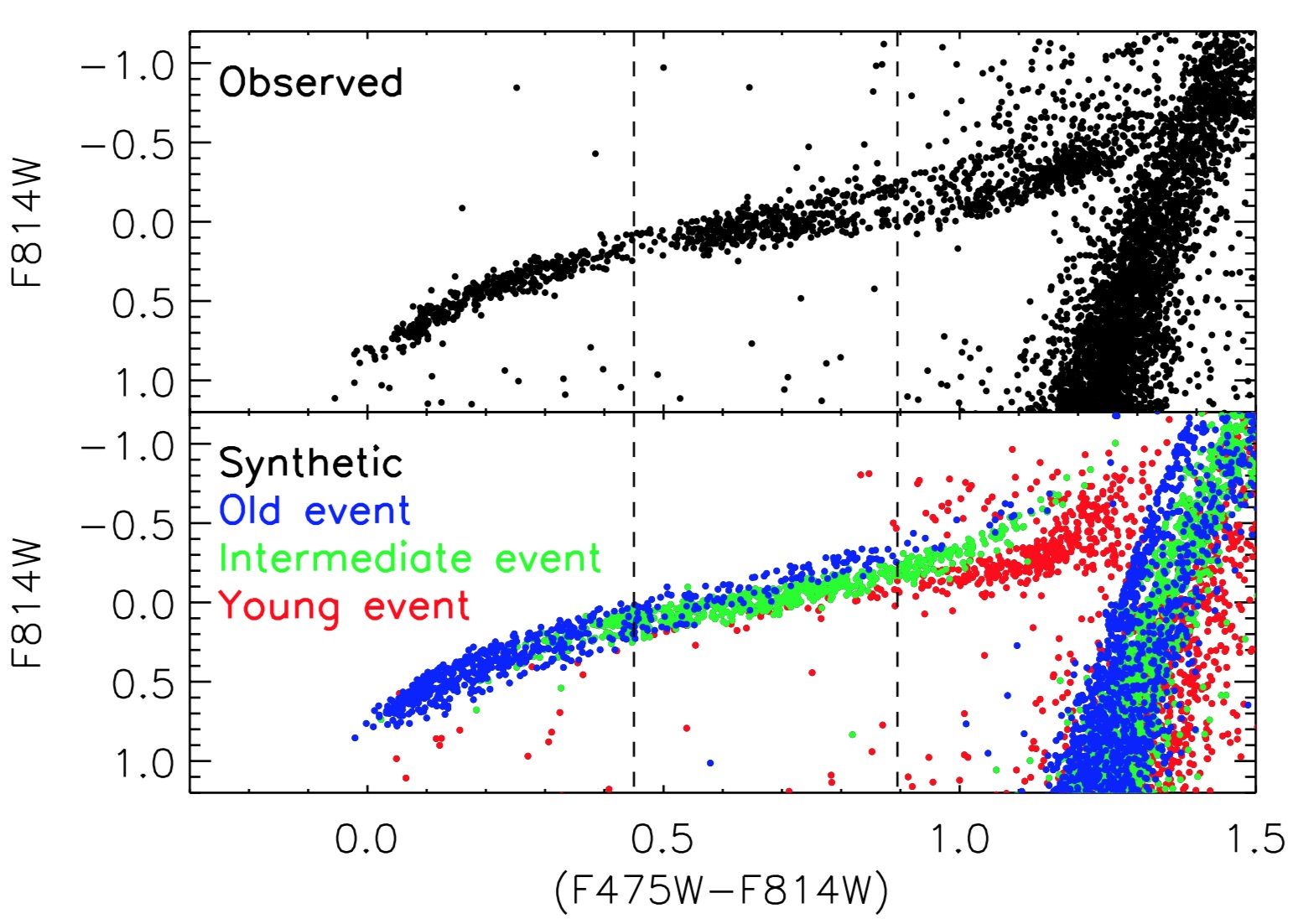}
\caption{Upper panel: the observed HB of Tucana. Lower panel: the synthetic HB  from our model. Stars belonging to the oldest star formation event are coloured in blue, those belonging to the intermediate event are coloured in green and stars belonging to the most recent event are coloured in red. The dashed lines mark the approximate position of the instability strip.}
\label{fig:burst}
\end{figure}

Our analysis of the morphology of the HB demonstrates that Tucana's SFH is composed of three distinct star formation events. These stellar populations correspond to the three stellar clumps observed on the HB. This is evident from Fig.~\ref{fig:burst}, that shows a comparison between the observed HB of Tucana and the synthetic HB coming from our model. In the synthetic CMD we identify the stars that belong to the three distinct episodes of star formation, and colour code them accordingly. Here we can clearly see the age-metallicity trend along the HB, discussed at the beginning of \S~\ref{redHB}. The position of the three star formation events on the HB matches that of the three observed stellar clumps, as expected.

In light of the results of \S~\ref{gaps}, the inability of our model SFH to reproduce the gaps on the HB, especially the bluer one, suggests that we are underestimating the burstiness of Tucana's SFH. This happens because our solution is limited by resolution effects, and it means that the two older events of star formation have a narrower distribution in age and metallicity than predicted by our model. This limitation of our resolution is determined in part by the photometric errors in the faint CMD regions. Of course, other effects can also contribute to the discrepancy, such as uncertainties related the colour excursions in the theoretical HB tracks and potential systematics in the reconstruction of the HB in the instability strip, where the gaps partially reside. However, although these effects can exacerbate the difference between our model and the observations, they are not the only cause.

Our analysis is not the first detection of distinct stellar populations in Tucana. Examining the luminosity function of the RGB, \citet{Monelli10c} reported the presence of two distinct RGB bumps. This was associated with the presence of two distinct stellar components, differing in metallicity. Another independent detection has been made, by the same group, looking at pulsational properties of the RR Lyrae \citep{Bernard08}. They divided the RR Lyrae into a ``bright" and a ``faint" sample, depending on their intensity-averaged magnitude, finding that the two samples defined different sequences in the pulsation period-amplitude diagram and suggesting a difference in metallicity between the two. These studies resulted in the conclusion that Tucana experienced two distinct events of star formation, very early on during its formation and separated by a short amount of time. This is confirmed by our model HB. From Fig.~\ref{fig:burst}, it can be seen that stars belonging to the intermediate event already enter the instability strip on the zero-age HB. Because of this, they are predominantly detected as low-luminosity RR Lyrae. In contrast, stars belonging to the oldest and most metal poor star formation event spend most of their life on the blue HB, and they cross the instability strip only when evolving towards the asymptotic giant branch. This difference in metallicity and evolutionary stage explains the higher luminosity of these stars.

The stars that formed during the final star formation event are confined to the reddest clump of the HB. These stars never cross the instability strip, and hence never become RR Lyrae. This is why no claims for a third star formation event in Tucana could be made by previous studies. However, as we have demonstrated, the presence of stars younger than previously thought is required to explain the HB morphology. Furthermore, the fact that our model underestimates the stellar counts on the red HB is an indication that the third star formation episode experienced by Tucana was stronger than we measure, possibly comparable with previous two.

The exact duration of the final star formation event, as we have shown, depends on the RGB mass loss at high metallicity. At values of [Fe/H] $\gtrsim -1.5$, our model predicts mass loss values of 0.15-0.20 $M_{\odot}$, resulting in the star formation lasting roughly 2 Gyr longer compared to M10. Of course, a lower mass loss value would soften the discrepancy between our determination and the SFH of M10. We currently assume a linear dependence of mass loss with metallicity, mostly due to computational constraints. Thus, any potential plateau of the mass loss value at the high-metallicity end would not be recovered by the present analysis. Such flattening at the high-metallicity end would also explain why we underpredict the stellar counts on the red HB clump, as too young models, necessary to reproduce the red HB with a linearly increasing mass loss, would be ruled out by the MSTO morphology. Clearly, a more sophisticated parametrization of the mass loss will need to be integrated in our procedure to disentangle these two quantities. Nevertheless, any non-negligible amount of RGB mass loss experienced by Tucana's stars still implies that the star formation persisted for some time after 8 Gyr ago.

 \begin{figure}
\centering
\includegraphics[width=0.5\textwidth]{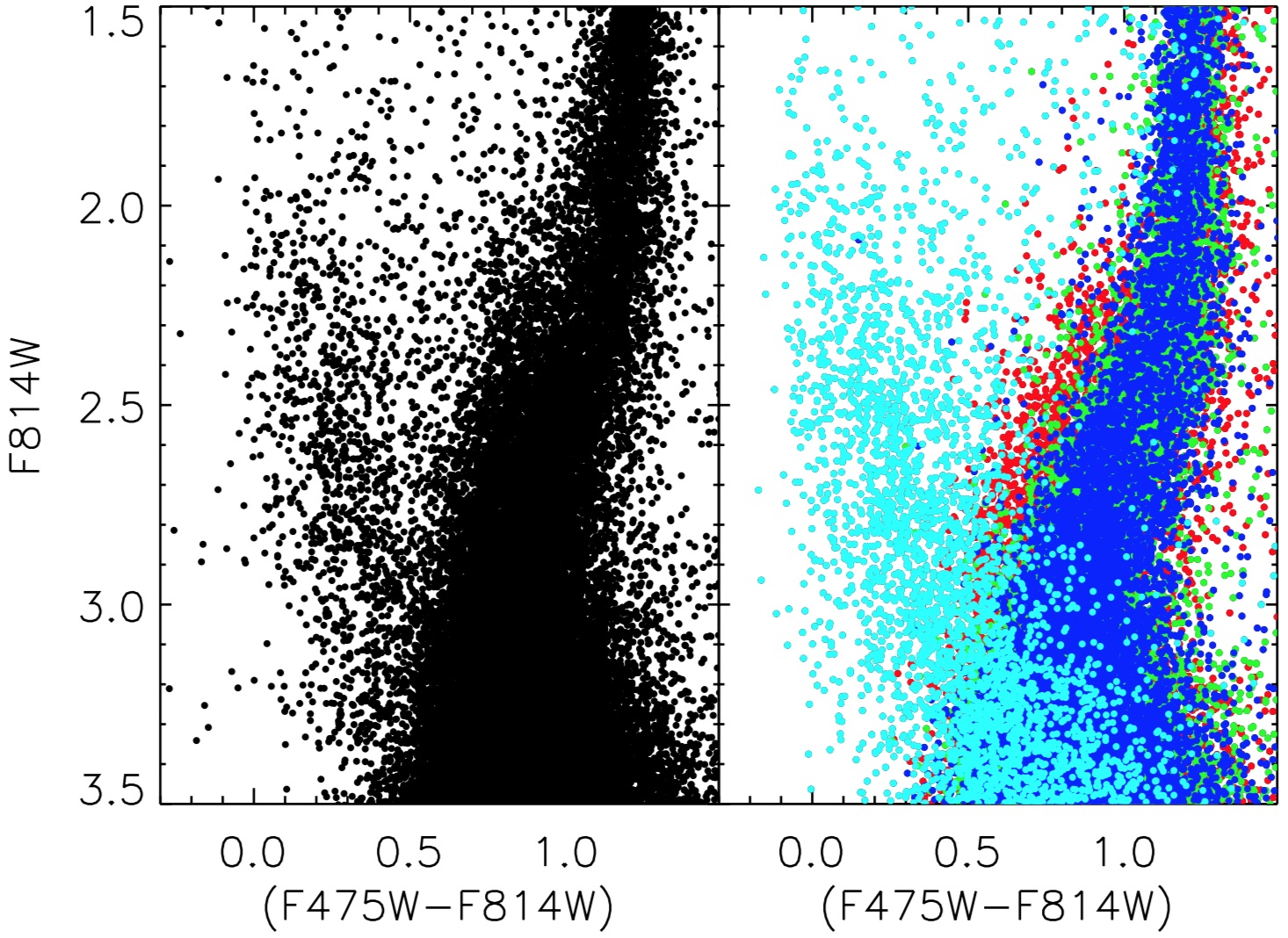}
\caption{Left panel: the observed MSTO of Tucana. Right panel: the synthetic MSTO  from our model. The colour scheme is the same of Fig.~\ref{fig:burst}, with the addition of the colour cyan for all the stars younger than 5.5 Gyr.}
\label{fig:MSTO}
\end{figure}

The multiplicity of Tucana's SFH could potentially manifest itself in the structure of the MSTO. For this reason we plot in Fig.~\ref{fig:MSTO}  the observed and synthetic MSTO region, similarly to Fig.~\ref{fig:burst}. This time, we make an additional distinction, identifying all the stars younger than 5.5 Gyr (cyan dots). This plot clearly shows that the blue plume stars are almost entirely confined in the sparse, very young, bins of our solution and thus they are unrelated to the last star formation event. Additionally, it is evident that the photometric uncertainties in this region prevent a clear separation of the star formation episodes. At this level of photometric precision, the old and intermediate population overlap entirely. The young population extends to slightly bluer colours on the subgiant branch and could tentatively be associated with what appears to be a small overdensity at $F814W \sim 2.2$ and $(F475W-F814W) \sim 0.8$. Interestingly, this feature seems to be similar to the ``spur'' observed above the MSTO of the Sculptor dSph \citep{Hurley-Keller99,Bettinelli19}.

\subsection{The spatial variation of Tucana's SFH}

\begin{figure}
        \subfloat[][]
	{\includegraphics[width=0.5\textwidth]{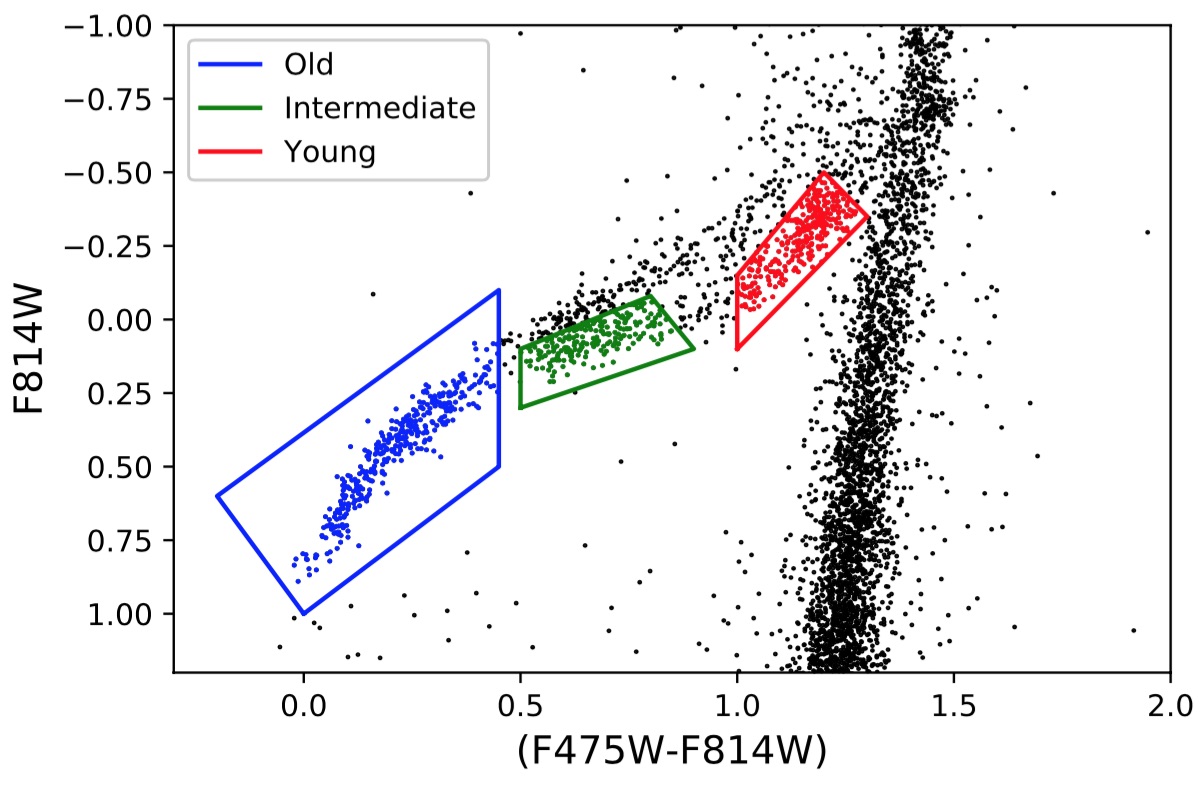} \label{fig:box}} \quad
	 \subfloat[][]
	{\includegraphics[width=0.5\textwidth]{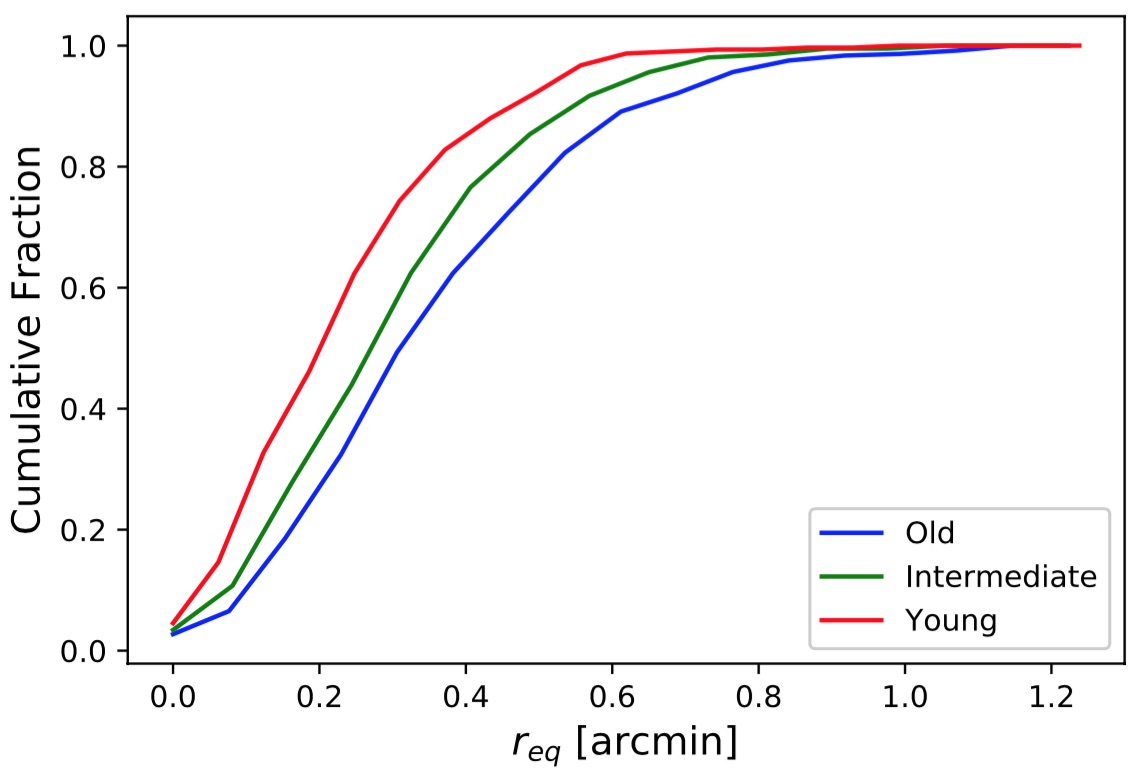} \label{fig:dist}} \quad
    \caption{a) Our selection criterion to isolate the three HB clumps of Tucana. The red and green box are designed to minimise contamination from less massive stars evolving towards the asymptotic giant branch. b) Cumulative radial distribution of the HB stars in the three clumps.}
    \label{fig:rad}
\end{figure}

The association between the features on the HB of dSph galaxies and the different phases of star formation activity makes it easy to determine the spatial distribution of their stellar sub-populations. The radial gradient in the HB morphology of Tucana was first reported by \citep{Harbeck01} who concluded it was sign of a metallicity gradient. However, the quality of the data available at the time allowed only to a comparison between red and blue HB, with the first one sensitive to contamination from the RGB. With the high photometric precision of our HST/ACS dataset, we can measure the spatial distribution of the three different HB clumps. This is shown in Fig.~\ref{fig:rad}. The coloured boxes in Fig.~\ref{fig:box} represent our selection for the old, intermediate and young age clumps. Care has to be taken when selecting stars in the intermediate and red HB clumps, as these regions are contaminated by lower mass HB stars that are evolving towards the asymptotic giant branch. These stars create the magnitude  bifurcation seen at $(F475W - F814W) > 0.8$, and if they are not removed properly, they will soften the true radial gradient. For this reason we selected stars in the intermediate and red HB clump close to the zero age HB, thus minimising the contamination from evolved stars. Our final selection consists of 367 stars in the blue clump, 206 in the intermediate clump and 308 stars in the red one.

The radial distribution of the three clumps is shown in Fig.~\ref{fig:dist}. As expected, we find a spatial gradient in the distribution of HB stars, which reflects a different spatial distribution of the distinct stellar generations. The older the population, the more diffuse is its spatial distribution. The younger stellar population, corresponding to the red HB clump is the most centrally concentrated. Such radial segregation is statistically significant. A two-sample Kolmogorov-Smirnov test shows that the young population has a probability of $1.69 \cdot 10^{-6}$ to be drawn from the spatial distribution of the intermediate population, and a probability of $1.89 \cdot 10^{-13}$ to have the same distribution of the old population. On the other hand, the old and the intermediate stellar populations have a 4\% probability to have the same radial distribution. Such stellar population gradient is typical of Local Group dSphs \citep[e.g.][]{Battaglia08, Vargas14} and, in general, of low-mass galaxies in the local Universe \citep[e.g.][]{Koleva11,Pilyugin15}. The simplest interpretation of these spatial gradients is that star formation in dwarf galaxies proceeded ``outside-in''. However, several studies have shown that both internal \citep[e.g.][]{El-Badry16} and external \citep[e.g.][]{Benitez-llambay16} mechanisms can induce a significant amount of stellar migration in low mass galaxies and can even reverse the original spatial gradient. In this framework, determining ages and radial profiles for the different subpopulations of dwarf galaxies can provide an interesting testing ground for such models.

 \section{Conclusions}
 \label{conclusions}

In this paper we carried out an accurate CMD modelling of the dwarf galaxy Tucana. Carrying on the work of \citet{Savino18a}, we confirmed how the simultaneous modelling of different features on the CMD is able to break the degeneracy between the SFH of a galaxy and the RGB mass loss experienced by its stars, providing precise measurements for both. The deep photometry provided by HST/ACS, with a reliable modelling of the associated photometric uncertainties, permitted to recover a self-consistent solution that satisfactorily reproduces the main features of the CMD.

The increased precision on the determination of the RGB mass loss provides useful constraints to theoretical models that aim to reproduce and characterise the mechanism beyond this poorly understood phenomenon. A comparison with previous measurements on the Sculptor dSph (Fig.~\ref{fig:MLscl}), shows good agreement within the uncertainties, suggesting that stars in different galaxies might experience similar amount of mass loss. Clearly, the modelling of deep CMDs from other external galaxies is required to ascertain whether a universal ``mass-loss law" exists for dSphs. If confirmed, such relation would be of great value for the interpretation of the bright CMD in distant galaxies. 

The simultaneous modelling of the HB and of the MSTO revealed that the SFH of Tucana is composed of three independent star formation events. The complexity of stellar populations in dSphs, and their spatial gradients, have been long known \citep[e.g.,][]{Harbeck01,Bellazzini01,Tolstoy04}. However, whether these galaxies host distinct stellar populations or a smooth gradient, in age or metallicity, is still unclear. Our result suggests that dSphs are composed by discrete stellar subpopulations, which can be clearly identified in the structure of the HB, demonstrating the strong link between the morphology of this CMD feature and the SFH of its galaxy. The radial distribution of Tucana's HB stars reveals a different concentration for the different star formation events, with younger stars preferentially found in the central regions of the galaxy.

The origin of these complex stellar populations is also debated. Scenarios for the formation of these galaxies include galaxy mergers \citep{Amorisco12a,delPino15}, in-situ star formation modulated by supernova feedback \citep{Salvadori08,Revaz09} or tidal interaction with larger galaxies \citep{Pasetto11}. At present, observational evidence is still not sufficient to provide a definitive answer. Thanks to the sophisticated modelling of the CMD we have been able, for the first time, to clearly separate these different subpopulation in the SFH and constrain their age and duration, making a step forward in our understanding of the early evolution of dwarf spheroidal galaxies.

\begin{acknowledgements}
We thank the anonymous referee for the useful comments that helped us improve the manuscript. We are grateful to the LCID group for providing us with the dataset used in this study. A.S. wishes to thank D. Massari and L. Posti for the useful discussions about this manuscript. This research has been supported by the Spanish Ministry
of Economy and Competitiveness (MINECO) under the grant 2017-89076-P.
\end{acknowledgements}

\bibliographystyle{aa}
\bibliography{./Bibliography}

\end{document}